\documentclass[prb,twocolumn,showpacs]{revtex4}
\usepackage{graphicx}
\begin{document}

\title{Inverse Magnetoresistance of Molecular Junctions}

\author{Hugh Dalgleish and George Kirczenow}

\affiliation{Department of Physics, Simon Fraser
University, Burnaby, British Columbia, Canada V5A 1S6}

\date{\today}

\begin{abstract}

We present calculations of spin-dependent electron
transport through single organic molecules bridging pairs
of iron nanocontacts. We predict the magnetoresistance of
these systems to switch from positive to negative with
increasing applied bias for both conducting and insulating
molecules. This inverse magnetoresistance phenomenon does
not depend on the presence of impurities and is unique to
nanoscale magnetic junctions.  Its physical origin is
identified and its relevance to experiment and to
potential technological applications is discussed.
\end{abstract} 


\maketitle

\section{Introduction}

Electronic devices composed of ferromagnetic electrodes
separated by a connecting junction whose resistance can be
altered with magnetic fields offer enormous potential for
technological applications such as magnetic random
access-memory (MRAM).  Spin-dependent transport in
ferromagnetic nanosystems is currently receiving increased
attention owing to the expectation that nanoscale size
constraints may maximize the potentially useful
magnetoresistance effects.  In particular, theories have
predicted that molecules bridging ferromagnetic electrodes
may present pathways for building ultra-small
spintronic devices that take advantage of weak molecular
spin-orbit interaction to maximize the
magnetoresistance.\cite{Emberly2002,Pati2003,Babiaczyk2004,
Rocha2005}  Recently, experiments have been reported in
which electrons transmitted through insulating octanethiol
molecular self-assembled monolayers (SAMs) between nickel
contacts\cite{Petta2004}, and through 100nm organic
molecular  films with magnetic
electrodes\cite{Dediu2002,Xiong2004},  retained their spin
polarization leading to observed magnetoresistance,
proving that individual or groups of organic molecules may
potentially be used as building blocks for the ultimate
miniaturization of spintronic devices.  However, bias
dependence and inversion of the magnetoresistance were
also reported and were attributed\cite{Petta2004} to the
presence of impurities within the molecular junction. 
Bias dependence of the magnetoresistance has since been
predicted in insulating molecular
junctions\cite{Rocha2005}, indicating that bias-dependent
magnetoresistance may be intrinsic to molecular spintronic
devices.  However, most technological device applications
based on magnetoresistance require a relatively
bias-independent magnetoresistance for stable device
operation. In this article, we report predictions of
magnetoresistance in ideal molecular spintronic junctions
where single molecules bridge iron electrodes, systems
that until now have received little experimental or
theoretical attention.  We predict strong bias dependence
in the positive magnetoresistance regime, but we also
predict the magnetoresistance to become negative with
increasing applied bias and we show that the
magnetoresistance is potentially larger and {\em stable}
under applied bias in this negative magnetoresistance
regime.  Our calculations reveal negative
magnetoresistance to be an {\em intrinsic} property of
some molecular junctions (i.e., unlike the effect reported
in Ni/octanethiol/Ni junctions\cite{Petta2004}, {\em not}
dependent on the presence of impurities) and thus offer a
path to stable operation of devices (such as molecular
MRAM) based on molecular spintronics but operating in the
negative magnetoresistance regime.  Therefore,
understanding the origins of negative magnetoresistance is
paramount to fundamental studies of spin-dependent
transport in molecular junctions and potentially
applicable to technological endeavors. 

We explore theoretically two very different molecular
junctions: a conducting (conjugated) benzene-dithiolate
(BDT) molecule and an insulating octane-dithiolate (ODT)
molecule bridging ferromagnetic Fe nanocontacts. The
phenomena that we predict appear in both.  

\section{Theory}

In our calculations the system is partitioned into
semi-infinite ideal ferromagnetic source and drain leads
and an extended molecular junction consisting of the
molecule and clusters of nearby Fe atoms\cite{cluster};
for BDT this ``extended molecule'' is shown  in the inset
of Fig.\ref{Fig1}(a).  The Fe-thiol bonding geometries are
estimated with {\em ab initio}
relaxations.\cite{G03}\cite{S-Fedist}   The electronic
structure of the Fe clusters is described by a
tight-binding Hamiltonian and non-orthogonal $s, p, d$
basis. The tight-binding parameters are based on fits to
{\em ab initio} band structures of Fe crystals\cite{Pap}
and have previously been employed successfully to study
magnetic multilayers\cite{Mathon2001},
magnetic\cite{Velev2004,Dalgleish2005} and
non-magnetic\cite{Cuevas1997} nanocontacts, and
ferromagnetic Ni molecular junctions\cite{Emberly2002}. 
For Fe structures with surfaces this model yields
spin-resolved surface densities of states and enhanced
surface magnetic moments\cite{Dalgleish2005} similar to those of
{\em ab initio} surface calculations\cite{Ohnishi1983}. 
Thus our model incorporates $both$ the bulk and surface
magnetic properties of Fe which together influence the
spin-dependent transport through a nanoscale junction of
bulk Fe leads bridged with molecules.  Spin flip processes
are not considered, consistent with weak spin-orbit
interaction in molecules and the high degree of spin
polarization retained in spin-dependent observations of
molecular junctions.\cite{Petta2004,Dediu2002,Xiong2004}
The molecular electronic parameters are described by a
tight-binding formalism based on extended-H{\"u}ckel (EH)
theory.\cite{EH}  This approach has been used successfully
to explain the experimental current-voltage
characteristics of molecular nanowires connecting
non-magnetic metal
electrodes.\cite{Datta1997,Emberly2001,Kushmerick2002} 
The EH parameters are based on atomic ionization energies
while the electronic parameters from Ref. \onlinecite{Pap}
describing the Fe clusters are defined up to an arbitrary
additive constant.  We adjust this constant to align the
Fermi energy of the contacts relative to the highest
occupied molecular orbital (HOMO) of the BDT molecule,
according to the difference in ionization energies of the
isolated structures estimated with {\em ab initio} total
energy calculations.  For Fe electrodes and the  molecules
considered here, this agrees well with the difference
between the work function of Fe and the HOMO energy of the
isolated molecule obtained from density functional theory,
a method that has been used successfully to align the
Fermi energy of gold with the HOMO of
BDT.\cite{Derosa2001} The potential profile of the
nanocontact arising from the bias voltage applied between
the two electrodes is difficult to calculate from first
principles since it is a non-equilibrium many-body
property. However, appropriate heuristic models for the
profile can yield accurate results for the
current.\cite{Ke2004} We adopt this approach here,
assuming the majority of the applied bias to drop over the
metal-molecule interface.\cite{Ke2004,Mulica2000,Damle2001}
\cite{potentialprofile} Nearly identical current-voltage
characteristics emerge if the details of the model
potential profile are varied. Thus these details are not
crucial to our predictions. Our transport
calculations are based on Landauer theory\cite{Dattasbook}
and Lippmann-Schwinger and Green's function techniques. 
We define the junction magnetoresistance as
$JMR=$
$\frac{(I_{par}-I_{anti})}{\frac{1}{2}(I_{par}+I_{anti})}$.
$I_{par}$ ($I_{anti}$) is the current through the junction
for parallel (anti-parallel) magnetizations on the Fe
electrodes.

\section{Results}

\begin{figure}
\includegraphics[width=0.99\columnwidth]{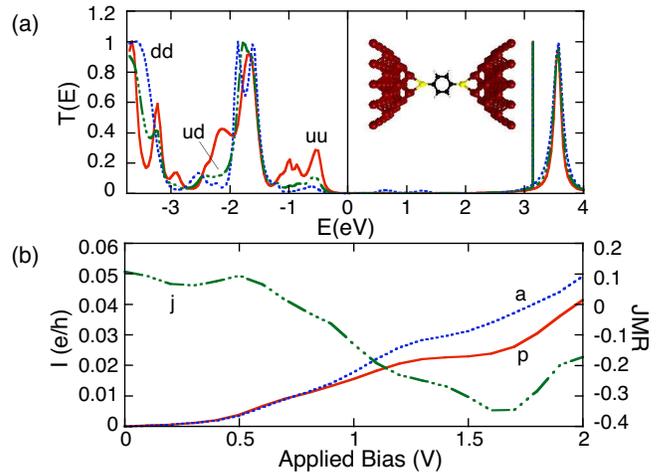}
\caption{(Color online) (a) Transmission probabilities vs.
energy at zero bias for the spin $up \rightarrow up$ (uu), 
$down \rightarrow down$ (dd), and $up \rightarrow down$ 
(ud) configurations of the BDT molecular junction shown 
in the inset.   Fermi energy = 0eV. (b) Current vs. 
voltage for parallel (p) and anti-parallel (a) 
magnetizations.  JMR (j) is also shown.}
\label{Fig1}
\end{figure}

Fig. \ref{Fig1}(a) shows the calculated transmission
probabilities at zero bias for the BDT molecular junction.
The spin
$up \rightarrow up$ (uu) and  $down \rightarrow down$ (dd)
transmission is for parallel magnetization of the two
contacts; $up \rightarrow down$ (ud) is for anti-parallel
magnetization. Quite similar results for the transmissions
and currents are  obtained if the 5x5 outer Fe layer is
deleted from each of the clusters indicating that the
clusters and ideal leads in our model adequately represent
real (macroscopic) Fe leads. The molecular HOMO and HOMO-1
give rise to the strong transmission 1.5-2eV below the
Fermi energy $E_F = 0$eV; the peaks 3-4eV above the Fermi
energy are due to the molecular LUMO and LUMO+1. 
The sharp resonance at 3.1eV results from extremely weak 
hybridization between the Fe leads and the isolated 
molecule's LUMO which has very little sulfur content.  
The transmission resonance is present and transmits for 
all spin configurations.  More
important, however, for potential device applications are
resonant states arising from strong hybridization between
the molecular sulfur and the Fe $d$-electron orbitals. In
Fig. \ref{Fig1}(a) these states give rise to the broad
transmission features between the Fermi energy and the
HOMO resonances mentioned above. They allow electrons to
transmit efficiently through the metal-molecule interface
and so give rise to moderately strong transmission
dominating the current in the experimentally accessible
moderate bias regime.  Together with the energetically
localized and highly spin-split Fe $d$ bands, these hybrid
states give rise to strongly spin-dependent transport
characteristics.  

\begin{figure}
\includegraphics[width=0.99\columnwidth]{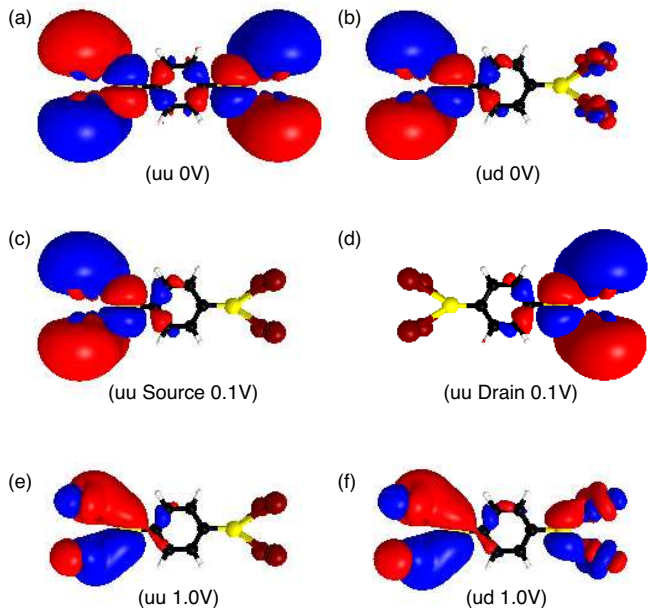}
\caption{(Color online) Representative hybridized energy
eigenstates of the entire extended molecule near the Fermi
energy are depicted on the BDT molecule and the 2x2 Fe
planes on either side.  (a) At 0V the uu source and drain
electrode $d$ states hybridize with sulfur orbitals on
$both$ sides of the molecule forming transmitting
states.   (b) Only the source strongly hybridizes with the
sulfur states in the ud configuration.  At 0.1V, uu states
in (a) split into a state mainly on the source (c) that
shifts upwards in energy and a state mainly on the drain
(d) that shifts downwards.  (e) At 1.0V uu hybridized
states on the source and drain have shifted well apart.
(f) Source
$up$ hybridized states are shifted into resonance with
drain $down$ states in the ud configuration creating
states that transmit strongly through the molecule.}
\label{Fig2}
\end{figure}

At zero applied bias for parallel magnetization of the
contacts, the spin $up$ Fe $d$ orbitals hybridize very
strongly with sulfur states on {\em both} sides of the
molecule as shown for a typical state in Fig.
\ref{Fig2}(a). This results in the dominant uu
transmission near (below) the Fermi energy in Fig.
\ref{Fig1}(a).  The spin
$down$ $d$ orbitals, occurring higher in energy, hybridize
only weakly with molecular sulfur states.  For
anti-parallel magnetizations [i.e., $up (down) \rightarrow
down (up)$ transmission], only the source (drain) contact
hybridizes strongly with the molecular end group (Fig.
\ref{Fig2}(b)), resulting in much weaker ud transmission
(anti-parallel magnetization) than the uu transmission in
Fig. \ref{Fig1}(a). As the hybrid states are not located 
{\em exactly} at the Fermi energy they contribute to the
current at very low bias through their off-resonant
tails.  

An applied bias breaks the symmetry between the source and
drain electrodes in the parallel magnetization
configuration. Thus the symmetric zero bias molecular
orbitals (as in Fig. \ref{Fig2}(a)) are supplanted by 
states associated mainly with either the source or drain 
electrode as shown in in Figure \ref{Fig2}(c,d) for the 
uu transmission.  The orbital in Figure
\ref{Fig2}(c) is shifted upwards in energy and is
associated with the source electrode, while that in Figure
\ref{Fig2}(d) is shifted downwards and is associated with
the drain.  These orbitals can be traced to the zero bias
orbital in Figure \ref{Fig2}(a) and their energy splitting
corresponds to the applied bias.  Since the energies of
these states are not perfectly aligned, simultaneous
resonant transmission through both ends of the molecule is
impossible.  However, since the states remain close in
energy at moderate bias, they
$both$ contribute $coherently$ to near-resonant
transmission via the extended molecule's Green's function
in the Lippmann-Schwinger scattering formalism: There is
appreciable coherent mixing between their off-resonant
transmission tails resulting in relatively strong uu
transmission. The corresponding molecular orbital for ud
transmission resembles that shown for the source electrode
in the uu configuration in Figure \ref{Fig2}(c), however a
state corresponding to the uu drain state is absent.  The
reverse is true for du transmission and since these two
transmission processes are mutually incoherent, there is
no mixing of their off-resonant tails and the current in
the case of anti-parallel magnetizations is weaker.  

Fig. \ref{Fig1}(b) shows the corresponding finite-bias
current-voltage characteristics.  At low bias, the current
grows as more hybridized states contribute to the
transmission.   Since the uu transmission dominates, the
parallel magnetization current is larger, resulting in a
positive magnetoresistance of about 10 percent. At larger
applied bias, the applied electrostatic potential breaks
the symmetry between the contacts further and the uu
transmission decreases as the hybridized Fe $d$ states
associated with the source and drain move apart.  However,
at high bias in the case of anti-parallel magnetizations,
hybridized states associated with $down$ $d$ electron
orbitals on the drain are shifted into resonance with $up$
$d$ electron states on the source and hybridized
transmitting states connecting {\em both} sides of the
molecule with the electrodes form in the case of $up
\rightarrow down$ transmission as shown in Fig.
\ref{Fig2}(f), enhancing the transmission.  No such
resonant effect occurs for du transmission which decreases
with increasing bias and so the {\em overall} effect of
the applied bias on the {\em net} transmission for
anti-parallel magnetization is small. No resonant
enhancement occurs at increased bias for parallel
magnetizations where hybridized states are shifted apart
(Fig. \ref{Fig2}(e)) and transmission is reduced.  
Therefore, because it selectively depresses  transmission 
for parallel magnetization of the contacts,  the 
bias-induced symmetry breaking results in a crossover 
with increasing bias from positive to negative values 
of the JMR in Fig. \ref{Fig1}(b).  

The wave functions of the hybridized states responsible
for transport at moderate bias are located primarily on
the Fe electrodes.  Therefore their energy levels are
pinned to the electrodes and shift rigidly with applied
bias explaining why our moderate bias results are
insensitive to the assumed potential profile. If one
partitions the entire system into halves associated with
either the source or with the drain so that the hybridized
states on the source (drain) side of the system are
grouped with the source (drain) to form a "composite"
source (drain) electrode, the negative magnetoresistance
argument can be rephrased in the language of conventional
solid-state magnetic tunnel junctions: At the bias
required to bring hybridized $down$ states on the drain
into resonance with hybridized $up$ states on the source,
the spin polarization of the composite drain is negative
(the current is dominated by $down$ electrons) while that
of  the composite source remains positive.  Therefore, the
Julliere \cite{Julliere1975} theory of tunneling would
predict negative magnetoresistance.

Because its mechanism relies on the simultaneous
suppression of transmission for parallel magnetizations on
the contacts and enhancement of the transmission in  the
anti-parallel ud transmission channel, this intrinsic
negative magnetoresistance requires strongly spin-split
energy bands providing a large magnetic moment.   No such
negative magnetoresistance has been reported for
calculations on molecules bridging ferromagnetic $nickel$
junctions
\cite{Emberly2002,Pati2003,Babiaczyk2004,Rocha2005} and our
calculations on similar systems suggest that this
different  behavior may be related to the smaller
spin-splitting in Ni: At sufficient bias to induce
resonant transmission in the ud channel (the required bias
is relatively small for Ni) much symmetry and overlap
remain in the uu transmission channel where energy levels
are not adequately shifted apart.  Therefore the mechanism
should be less effective in nanosystems based on nickel
\cite{Nineg}\cite{Mn}.

For BDT bridging {\em Fe} nanocontacts, the crossover to
negative magnetoresistance occurs at relatively low bias,
near 0.7V, well within the range of current experimental
techniques and therefore the predicted negative
magnetoresistance should be accessible to experiment. 
With magnitude reaching a value of more than 30 percent,
the negative magnetoresistance may be more easily observed
in this system than positive values.   Clearly the
mechanism is expected to be unique to systems that can
both sustain a significant potential drop and allow for
resonant transmission of $d$-electrons through the
junction, i.e., nanosystems.  Negative magnetoresistance
emerging as a result of applied bias has already been
observed experimentally in systems where ferromagnetic
electrodes are separated by thin film tunnel junctions of
only a few nanometers in width where the junction couples
strongly to the electrode $d$ states
\cite{Sharma1999,Teresa1999}. Since $d$-electrons are not
expected to contribute to transport in conventional thin
film magnetic tunnel
junctions\cite{Mathon2001,MacLaren1997,
Butler2001,Tsymbal2003}, negative magnetoresistance when
observed in those devices is attributed to
impurities.\cite{TsymbalSokolov2003}  However, we predict
{\em intrinsic} negative magnetoresistance in ideal
systems demonstrating that bias-dependent and negative
magnetoresistance are not necessarily signatures of
impurities in molecular nanosystems.

Interestingly, the region exhibiting the greatest bias
independence in our magnetoresistance calculations emerges
in the negative magnetoresistance regime (around 1.7V in
Figure \ref{Fig1}(b)).  Since magnetoresistance devices
can in principle exploit magnetoresistance of either sign
but require a relatively bias independent
magnetoresistance for stable device operation,
molecular-Fe systems may be attractive candidates for
technological applications in devices based on molecular
spintronics (molecular MRAM) but operated in the negative
magnetoresistance regime. 

For comparison, we have also studied an Fe/BDT/Fe
molecular junction having a different geometry where
single Fe tip atoms have been added to each of the
clusters and the molecular sulfur atoms bind directly over
them at the same heights considered previously. This
geometry results in stronger and sharper resonances
associated with the hybridized states and therefore the
mechanism responsible for negative JMR is even more
prominent.   Here we find the cross-over to  negative
magnetoresistance to occur at a lower bias voltage of
$0.25V$. 

\begin{figure}
\includegraphics[width=0.99\columnwidth]{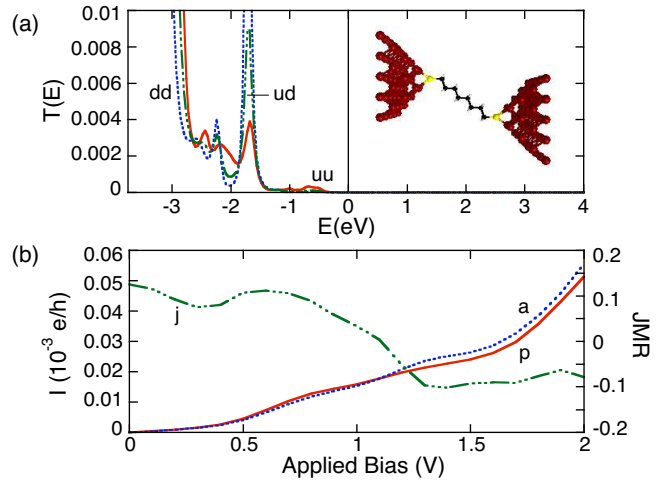}
\caption{(Color online) Calculated spin-dependent
transmission, currents and JMR of the ODT molecular
junction shown in the inset. Notation as in
Fig.\ref{Fig1}.}
\label{Fig3}
\end{figure}

Next we consider ferromagnetic Fe contacts bridged with an
octane-dithiolate (ODT) molecule, shown in Fig.
\ref{Fig3}(a) with the zero bias transmission
probabilities.  As ODT is an insulator, the strongest
transmission features near the Fermi energy are weak
resonant features associated with the HOMO, about 1.8eV
below the Fermi energy.  Other weakly transmitting states
appear closer to the Fermi energy (about 0.5 eV below) due
to hybridization of Fe
$d$ and molecular sulfur orbitals.  These states are
similar to those predicted in the case of BDT: the
$up \rightarrow up$ transmission is largest as
hybridization occurs at both ends of the molecule allowing
for efficient transmission into and out of the
metal-molecule interface, decreasing the width of the
tunnel gap.  The calculated current-voltage
characteristics  are shown in Figure
\ref{Fig3}(b).  At moderate bias, the symmetry between the
source and drain electrodes in the parallel magnetization
configuration leads to positive magnetoresistance.  At
higher applied bias, by the same mechanism described for
BDT, the magnetoresistance becomes negative, but here the
crossover doesn't occur until nearly 1.2V.  Again a large
region of relative bias stability emerges in the negative
bias regime (around 1.5V in Figure
\ref{Fig3}(b)) with possible implications for future
molecular spintronic technology.

\section{Conclusion}

In conclusion: We have predicted an inverse
magnetoresistance phenomenon $intrinsic$ to systems where
ferromagnetic Fe electrodes are bridged by two very
different molecules (conducting and insulating), and for
different bonding geometries. We also predict
qualitatively similar behavior in Fe junctions bridged by
a pair of Fe atoms.\cite{Dalgleish2005} Thus negative
magnetoresistance should be available in a variety of
molecular junctions under appropriate conditions.  The
value of the bias required to induce negative
magnetoresistance depends on the chemistry of the
junction: 0.25-0.7V for Fe/benzene-dithiolate/Fe
(depending on the bonding geometry), 1.2V for
Fe/octane-dithiolate/Fe and higher still\cite{Dalgleish2005} 
for Fe contacts bridged by a Fe dimer. These changes in
the chemistry make it possible to vary the onset of
negative magnetoresistance, the magnitudes of both
positive and negative magnetoresistances and the regions
of bias independence important for device applications.   

\section{Acknowledgement}

This work was supported by NSERC and the Canadian
Institute for Advanced Research.


\begin{thebibliography}{}
\bibitem{cluster} Each Fe cluster of the extended molecule
is built from 5x5, 4x4, 3x3, 2x2 (100)-oriented layers (54
atoms total) of atoms  in the bulk geometry of bcc iron.
\bibitem{potentialprofile} We assume that one-third of the
applied bias drops at each interface and the remaining
one-third drops linearly over the length the molecule.
\end{thebibliography}
\end{document}